\documentclass[sigplan,nonacm,10pt,screen]{acmart}
\renewcommand\footnotetextcopyrightpermission[1]{}
\usepackage[normalem]{ulem}

\usepackage{hyperref}
\hypersetup{hidelinks,colorlinks=true,linkcolor=red,citecolor=purple}

\usepackage{soul}
\usepackage{enumitem}

\usepackage{setspace}
\usepackage{epsfig}
\usepackage{graphicx}
\usepackage{subcaption}
\usepackage{afterpage}
\usepackage{times}
\usepackage{amsmath}

\usepackage{bbding}
\usepackage{pifont}

\usepackage{amssymb}
\usepackage{wasysym}
\usepackage{latexsym}

\usepackage{url}

\usepackage{multirow}

\renewenvironment{thebibliography}[1]{%
\begin{oldthebibliography}{#1}%
  \setlength{\parskip}{0ex}%
  \setlength{\itemsep}{0ex}%
}%
{%
\end{oldthebibliography}%
}

\usepackage{listings}
  \usepackage{courier}
\usepackage[textsize=tiny,textwidth=0.6in]{todonotes}
\newcommand{\allnotes}[1]{}
\renewcommand{\allnotes}[1]{#1} 

\renewcommand{\em}{\it}

\newcommand{\ignore}[1]{}

\def\cfigure[#1,#2,#3]{
\begin{figure}
\vspace*{0mm}
\begin{center}

\includegraphics[width=3in]{#1} 
 
\vspace*{-3mm}\caption[]{#2
} \label{#3}
 
\vspace*{-5mm}
\end{center}
\end{figure}}

\def\cfigurefour[#1,#2,#3]{
\begin{figure}
\vspace*{0mm}
\begin{center}

\includegraphics[width=4in]{#1} 
 
\vspace*{-3mm}\caption[]{#2
} \label{#3}
 
\vspace*{-5mm}
\end{center}
\end{figure}}

\def\cfiguretemp[#1,#2,#3]{
\begin{figure}
\vspace*{0mm}
\begin{center}

\includegraphics[width=3.5in]{#1} 
 
\vspace*{-3mm}\caption[]{#2
} \label{#3}
 
\vspace*{-5mm}
\end{center}
\vspace*{-2mm}
\end{figure}}

\def\wfigure[#1,#2,#3]{
\begin{figure*}
\vspace*{0mm}
\begin{center}
 \includegraphics[width=\textwidth]{#1} 
 \vspace*{-3mm}\caption[]{#2
} \label{#3}
 
\end{center}
\end{figure*}}

\def\threefigure[#1,#2,#3,#4,#5]{
\begin{figure*}
\vspace*{0mm}
\begin{center}

\begin{tabular}{ccc}
\includegraphics[width=2in]{#1} & \includegraphics[width=2in]{#2} &  \includegraphics[width=2in]{#3} \\
(a) & (b) & (c) \\
\end{tabular}

\vspace*{-3mm}\caption[]{#4
} \label{#5}

\vspace*{-5mm}
\end{center}
\vspace*{-2mm}
\end{figure*}}

\def\dcfigure[#1,#2,#3,#4,#5,#6]{
{
\begin{figure*}
\begin{center}
\begin{minipage}[c]{\columnwidth}{
\includegraphics[width=\columnwidth]{#1} 
\vspace*{0mm}\caption[]{#2} \label{#3} \
}\end{minipage}\hspace*{\columnsep}\
\begin{minipage}[c]{\columnwidth}{
\includegraphics[width=\columnwidth]{#4} 
\vspace*{0mm}\caption[]{#5}\label{#6} \
}\end{minipage}
\end{center}
\end{figure*}
}
}

\def\tableByTable[#1,#2,#3,#4,#5,#6]{
{
\begin{table*}
\begin{center}
\begin{minipage}[c]{3in}{
\centering
{#1}
\vspace*{0mm}\tabcaption[]{#2}\label{#3} \
}\end{minipage}\hspace*{\columnsep}\
\begin{minipage}[c]{3in}{
\centering
{#4}
\vspace*{0mm}\tabcaption[]{#5}\label{#6} \
}\end{minipage}
\end{center}
\end{table*}
}
}

\def\figureByTable[#1,#2,#3,#4,#5,#6]{
{
\begin{figure*}
\begin{center}
\begin{minipage}[c]{3in}{
\centering
\includegraphics[width=\textwidth]{#1}
\vspace*{0mm}\figcaption[]{#2} \label{#3} \
}\end{minipage}\hspace*{\columnsep}\
\begin{minipage}[c]{3.3in}{
\centering
{#4}
\vspace*{0mm}\tabcaption[]{#5}\label{#6} \
}\end{minipage}
\end{center}
\end{figure*}
}
}

\def\tableByFigure[#1,#2,#3,#4,#5,#6]{
{
\begin{figure*}
\begin{center}
\begin{minipage}[c]{4.3in}{
\centering
{#1}
\vspace*{0mm}\tabcaption[]{#2} \label{#3} \
}\end{minipage}\hspace*{\columnsep}\
\begin{minipage}[c]{2.2in}{
\centering
\includegraphics[width=\textwidth]{#4}
\vspace*{-0.35in}\caption[]{#5}\label{#6} \
}\end{minipage}
\end{center}
\end{figure*}
}
}

\def\doublecfigure[#1,#2,#3,#4]{
{
\begin{figure}
\begin{center}
\begin{minipage}[c]{1.5in}{
\begin{center}
\includegraphics[width=1.5in]{#1}
\end{center}
}\end{minipage}\hspace*{1em}\
\begin{minipage}[c]{1.5in}{
\begin{center}
\includegraphics[width=1.5in]{#2}
\end{center}
}\end{minipage}
\vspace*{0mm}\caption[]{#3} \label{#4} \
\end{center}
\end{figure}
}
}

\def\qcfigure[#1,#2,#3,#4,#5,#6]{
{
\begin{figure*}
\vspace*{0.2in}\
\begin{center}
\begin{minipage}[c]{3in}{
\includegraphics[width=3in]{#1} 
\vspace*{-3mm}
}
\end{minipage}\hspace*{0.5in}\
\begin{minipage}[c]{3in}{
\includegraphics[width=3in]{#2} 
\vspace*{-3mm}
}\end{minipage}

\begin{minipage}[c]{3in}{
\includegraphics[width=3in]{#3} 
\vspace*{-3mm}
}
\end{minipage}\hspace*{0.5in}\
\begin{minipage}[c]{3in}{
\includegraphics[width=3in]{#4} 
\vspace*{-3mm}
}\end{minipage}
\end{center}
\caption[]{#5}\label{#6}
\end{figure*}
}
}

\def\twfigure[#1,#2,#3,#4,#5]{
{
\begin{figure*}
\vspace*{0.2in}\
\begin{center}
\begin{minipage}[c]{6.5in}{
\includegraphics[width=6.5in]{#1} 
\vspace*{-3mm}
}
\end{minipage}

\begin{minipage}[c]{6.5in}{
\includegraphics[width=6.5in]{#2} 
\vspace*{-3mm}
}\end{minipage}

\begin{minipage}[c]{6.5in}{
\includegraphics[width=6.5in]{#3} 
\vspace*{-3mm}
}
\end{minipage}
\end{center}
\caption[]{#4}\label{#5}
\end{figure*}
}
}

\def\dwfigure[#1,#2,#3,#4]{
{
\begin{figure*}
\vspace*{0.2in}\
\begin{center}
\begin{minipage}[c]{6.5in}{
\includegraphics[width=6.5in]{#1} 
\vspace*{-3mm}
}
\end{minipage}

\begin{minipage}[c]{6.5in}{
\includegraphics[width=6.5in]{#2} 
\vspace*{-3mm}
}\end{minipage}

\end{center}
\caption[]{#3}\label{#4}
\end{figure*}
}
}

\def\dssfigure[#1,#2,#3,#4,#5,#6]{
{
\begin{figure*}
\vspace*{0.2in}\
\begin{center}
\begin{minipage}[c]{4in}{
\includegraphics[width=4in]{#1}
\vspace*{-3mm}\caption[]{#2} \label{#3} \
}\end{minipage}\hspace*{0.5in}\
\begin{minipage}[c]{2in}{
\includegraphics[width=2in]{#4}
\vspace*{-3mm}\caption[]{#5}\label{#6} \
}\end{minipage}
\end{center}
\vspace*{-0.4in}\
\end{figure*}
}
}

\def\dsfigure[#1,#2,#3,#4,#5,#6]{
{
\begin{figure*}
\vspace*{0.2in}\
\begin{center}
\begin{minipage}[c]{3in}{
\includegraphics[width=3in]{#1}
\vspace*{-3mm}\caption[]{#2} \label{#3} \
}\end{minipage}\hspace*{0.5in}\
\begin{minipage}[c]{3in}{
\hspace*{0.5in}\
\includegraphics[height=3in]{#4}
\vspace*{-3mm}\caption[]{#5}\label{#6} \
}\end{minipage}
\end{center}
\vspace*{-0.4in}\
\end{figure*}
}
}

\def\dsyfigure[#1,#2,#3,#4,#5,#6]{
{
\begin{figure*}
\vspace*{0.2in}\
\begin{center}
\begin{minipage}[c]{2.5in}{
\includegraphics[height=2.5in]{#1}
\vspace*{-3mm}\caption[]{#2} \label{#3} \
}\end{minipage}\hspace*{0.5in}\
\begin{minipage}[c]{2.5in}{
\includegraphics[height=2.5in]{#4}
\vspace*{-3mm}\caption[]{#5}\label{#6} \
}\end{minipage}
\end{center}
\vspace*{-0.4in}\
\end{figure*}
}
}

\def\dyfigure[#1,#2,#3,#4,#5,#6]{
{
\begin{figure*}
\vspace*{0.2in}\
\begin{center}
\begin{minipage}[c]{3in}{
\includegraphics[height=3in]{#1} 
\vspace*{-3mm}\caption[]{#2} \label{#3} \
}\end{minipage}\hspace*{0.5in}\
\begin{minipage}[c]{3in}{
\includegraphics[height=3in]{#4} 
\vspace*{-3mm}\caption[]{#5}\label{#6} \
}\end{minipage}
\end{center}
\vspace*{-0.4in}\
\end{figure*}
}
}

\def\dyoldfigure[#1,#2,#3,#4,#5,#6]{
{
\begin{figure*}
\vspace*{0.2in}\
\begin{center}
\begin{minipage}[c]{3in}{
\epsfysize=2.0in\
\hspace{0.5in}\
\epsfbox{#1}
\vspace*{-3mm}\caption[]{#2} \label{#3} \
}\end{minipage}\hspace*{0.25in}\
\begin{minipage}[c]{3in}{
\epsfysize=2.0in\
\hspace{0.5in}\
\epsfbox{#4}
\vspace*{-3mm}\caption[]{#5}\label{#6} \
}\end{minipage}
\end{center}
\vspace*{-0.4in}\
\end{figure*}
}
}

\def\cfiguredouble[#1,#2,#3,#4]{
\begin{figure}
\vspace*{0.2in}\
\begin{center}
\begin{minipage}[c]{1.5in}{
\epsfxsize=1.5in\
\epsfbox{#1}
}\end{minipage}\hspace*{0.1in}\
\begin{minipage}[c]{1.5in}{
\epsfxsize=1.5in\
\vspace{0.1in}\epsfbox{#2}
}\end{minipage}\vspace*{-0.10in} \caption[]{#3}\label{#4}
\end{center}
\vspace*{-0.4in}\
\end{figure}
}

\def\wpfigure[#1,#2,#3,#4]{
\begin{figure*}
\vspace*{4mm}
\begin{center}

\includegraphics[width=#4]{#1} 

\vspace*{-3mm}\caption[]{#2
} \label{#3}

\vspace*{-5mm}
\end{center}
\end{figure*}}

\def\wprfigure[#1,#2,#3,#4,#5]{
\begin{figure*}
\vspace*{4mm}
\begin{center}

\includegraphics[width=#4, angle=#5]{#1} 

\vspace*{-3mm}\caption[]{#2
} \label{#3}

\vspace*{-5mm}
\end{center}
\end{figure*}}

\def\DoubleFigureWSlide[#1,#2,#3,#4,#5,#6,#7,#8,#9]{
\begin{figure*}
\vspace*{#9}
\begin{center}
\begin{minipage}{#4}
\includegraphics[width=#4]{#1}
\vspace*{-3mm}\caption{#2
}\label{#3}
\end{minipage}
\hspace{2em}
\begin{minipage}{#8}
\includegraphics[width=#8]{#5}
\vspace*{-3mm}\caption{#6
}\label{#7}
\end{minipage}
\vspace*{-5mm}
\end{center}
\end{figure*}
}

\def\DoubleFigureW[#1,#2,#3,#4,#5,#6,#7,#8]{
\begin{figure*}
\vspace*{0in}
\begin{center}
\begin{minipage}{#4}
\includegraphics[width=#4]{#1}
\vspace*{-3mm}\caption{#2
}\label{#3}
\end{minipage}
\hspace{2em}
\begin{minipage}{#8}
\includegraphics[width=#8]{#5}
\vspace*{-3mm}\caption{#6
}\label{#7}
\end{minipage}
\vspace*{-5mm}
\end{center}
\end{figure*}
}

\def\DoubleFigureWHack[#1,#2,#3,#4,#5,#6,#7,#8]{
\begin{figure*}
\vspace*{0in}
\begin{center}
\begin{minipage}{3in}
\includegraphics[width=#4]{#1}
\vspace*{-3mm}\caption{#2
}\label{#3}
\end{minipage}
\hspace{2em}
\begin{minipage}{3in}
\includegraphics[width=#8]{#5}
\vspace*{-3mm}\caption{#6
}\label{#7}
\end{minipage}
\vspace*{-5mm}
\end{center}
\end{figure*}
}

\def\ddcfigure[#1,#2,#3,#4]{
\begin{figure*}
\vspace*{0.2in}\
\begin{center}
\begin{minipage}[c]{\columnwidth}{
\includegraphics[width=\columnwidth]{#1} 
}\end{minipage}\hspace{0.5in}\
\begin{minipage}[c]{\columnwidth}{
\includegraphics[width=\columnwidth]{#2} 
}\end{minipage} \caption[]{#3}\label{#4}
\end{center}
\end{figure*}
}

\def\ddcfigureSlide[#1,#2,#3,#4,#5]{
\begin{figure*}
\vspace*{#5}\
\begin{center}
\begin{minipage}[c]{3in}{
\includegraphics[height=3in]{#1} 
}\end{minipage}\hspace{0.5in}\
\begin{minipage}[c]{3in}{
\includegraphics[height=3in]{#2} 
}\end{minipage}\vspace*{-0.10in} \caption[]{#3}\label{#4}
\end{center}
\vspace*{-0.4in}\
\end{figure*}
}

\def\cxfigure[#1,#2,#3]{
\begin{figure}
\vspace*{4mm}
\begin{center}
 
\epsfxsize=2.5in\
\epsfbox{#1}\
 
\vspace*{-0.10in}\caption[]{#2
} \label{#3}
 
\vspace*{-5mm}
\end{center}
\vspace*{-2mm}
\end{figure}}

\newcommand{\beforecaption}{\vspace{-.15cm}\begin{spacing}{0.85}}
\newcommand{\aftercaption}{\vspace{-.45cm}\end{spacing}}

\newcommand{\eg}{\textit{e.g.}}
\newcommand{\ie}{\textit{i.e.}}

\newcommand{\boldpara}[1]{\noindent{\textbf{#1}}}

\newcommand{\boldunderpara}[1]{\noindent{\underline{\textbf{#1}}}}

\newcommand{\projectname}{VDCores}

\newcommand{\sys}{\projectname}

\usepackage{tikz}

\newif\ifremark
\long\def\remark#1{
\ifremark%
        \begingroup%
        \dimen0=\columnwidth
        \advance\dimen0 by -1in%
        \setbox0=\hbox{\parbox[b]{\dimen0}{\protect\em #1}}
        \dimen1=\ht0\advance\dimen1 by 2pt%
        \dimen2=\dp0\advance\dimen2 by 2pt%
        \vskip 0.25pt%
        \hbox to \columnwidth{%
                \vrule height\dimen1 width 3pt depth\dimen2%
                \hss\copy0\hss%
                \vrule height\dimen1 width 3pt depth\dimen2%
        }%
        \endgroup%
\fi}

\newcommand{\mop}{$\mu$op}
\newcommand{\mops}{$\mu$ops}
\newcommand{\blackcircled}[1]{\tikz[baseline=(char.base)]{\node[shape=circle,fill=black,text=white,inner sep=1pt] (char) {#1};}}
\newcommand{\uop}{\mop}

\begin{document}

\title[]{\sys: Resource Decoupled Programming and Execution for Asynchronous GPUs}

\author{Zijian He}
\email{zih015@ucsd.edu}
\affiliation{%
  \institution{University of California, San Diego}
  \city{San Diego}
  \state{California}
  \country{USA}
}

\author{Adrian Sampson}
\email{asampson@cs.cornell.edu}
\affiliation{%
  \institution{Cornell University}
  \city{Ithaca}
  \state{New York}
  \country{USA}
}

\author{Yiying Zhang}
\email{yiying@ucsd.edu}
\affiliation{%
  \institution{University of California, San Diego and GenseeAI Inc.}
  \city{San Diego}
  \state{California}
  \country{USA}
}

\author{Zhiyuan Guo}
\email{zhiyuang@cornell.edu}
\affiliation{%
  \institution{Cornell University}
  \city{Ithaca}
  \state{New York}
  \country{USA}
}

\settopmatter{printfolios=true, printacmref=false}

\begin{abstract}

Modern GPUs increasingly rely on specialized and asynchronous hardware units to deliver high performance. 
Yet these units are often underutilized because today’s GPU software stacks still organize programming and execution around a monolithic kernel model that mismatches asynchronous hardware.
To address this issue,
Virtual Decoupled Engines (\sys{}) presents a new decoupled programming and execution model for asynchronous GPUs. \sys{} abstracts asynchronous hardware execution units as resource isolated virtual cores and represents workloads as dependency-connected micro-operations (\mops{}).
this abstraction removes static orchestration from the programmer,
enables automatic overlap of memory and compute based on dependency and resource readiness,
and thereby improves utilization of asynchronous hardware resources.

Realizing such a decoupled abstraction efficiently on today’s GPUs is itself challenging, \sys{} addresses this through a GPU-specialized programming model and GPU runtime design that preserves the flexibility while minimizing implementation overhead.
Across four LLM inference workloads on GH200, H100, and RTX 6000 Pro GPUs, \sys{} significantly improves decoding throughput by 24\% on average and by up to 77\% under dynamic inputs, while reducing kernel programming and specialization effort by 90\%.
We have open sourced \sys{} at \url{https://github.com/vdcores/vdcores}.

\end{abstract}

\maketitle

\section{Introduction}
\label{sec:intro}

\begin{figure*}[t]
\begin{minipage}[t]{0.66\textwidth}
\centering
\includegraphics[width=0.96\textwidth]{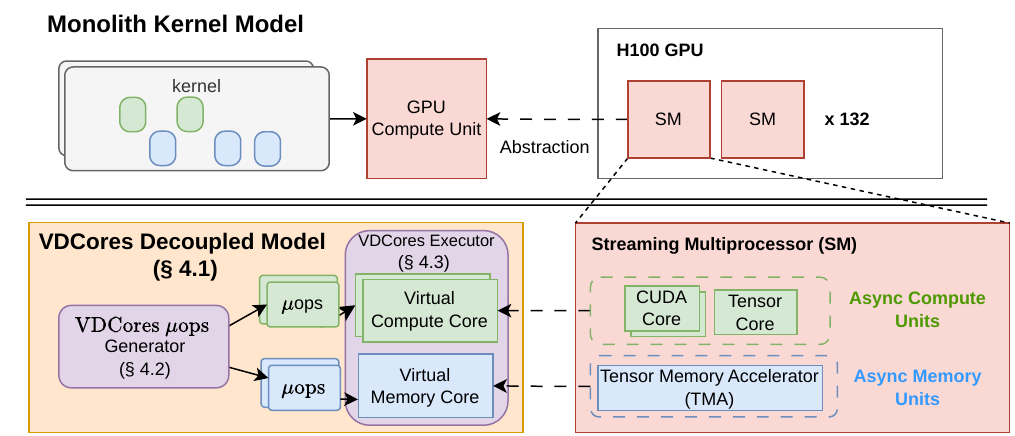}
\caption{Comparing \sys{} and monolith kernel programming and execution models.~\footnotesize{On an NVIDIA H100 GPU with asynchronous hardware units.}}
\label{fig:vdcores-programming-model}
\end{minipage}
\hfill
\begin{minipage}[t]{0.31\textwidth}
\centering
\includegraphics[width=0.96\textwidth]{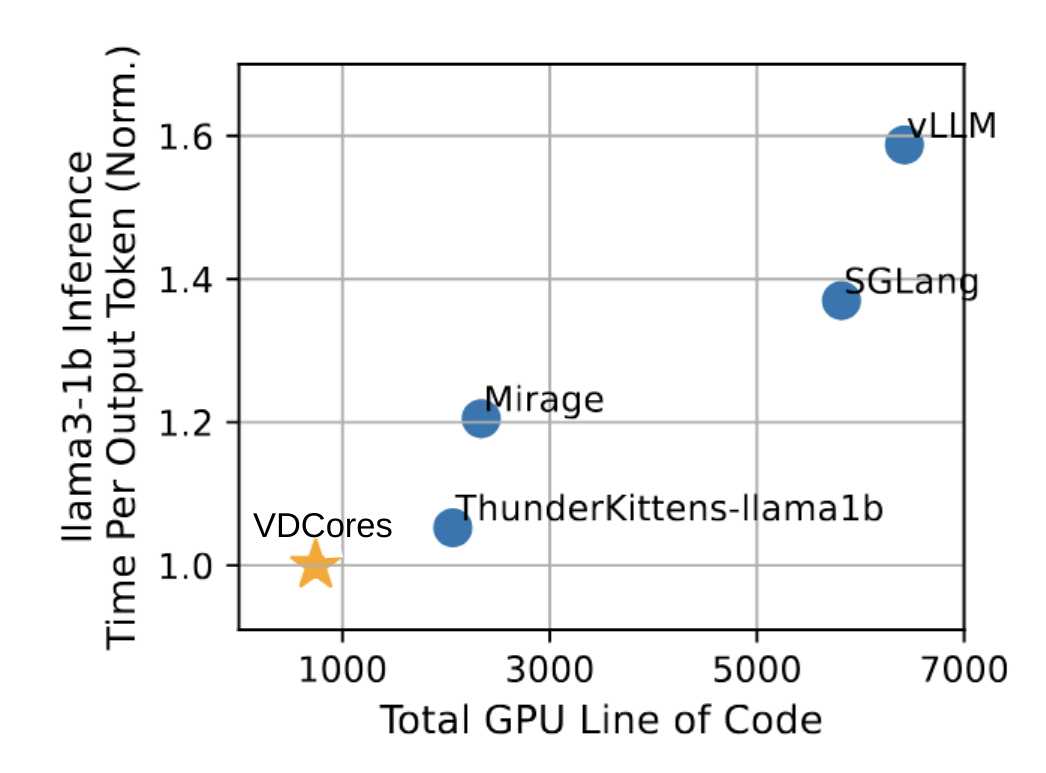}
\caption{\sys{} Inference Performance and Programming Effort. \textbf{Bottom left is better}.}
\label{fig:perf-intro}
\end{minipage}
\end{figure*}

Modern GPUs increasingly expose specialized and asynchronous internal resources, such as tensor cores and hardware-assisted asynchronous memory movement~\cite{nvidia_tensor_cores,nvidia_cuda_async}. Architectures such as NVIDIA Hopper further strengthen this trend with mechanisms like the Tensor Memory Accelerator (TMA), allowing memory transfer and computation to proceed concurrently~\cite{nvidia_hopper_blog}. In principle, such hardware should improve utilization by reducing idle time and enabling overlap across different execution units.

However, GPU programming and execution models have not evolved accordingly.
Most existing systems still rely on a \emph{resource-coupled monolithic kernel model}, originally designed for largely synchronous, data-parallel execution.
They continue to use this model even as GPU hardware becomes increasingly asynchronous and task-parallel.
This model fundamentally misaligns with modern GPU hardware and creates two drawbacks.

First, the monolithic kernel abstraction substantially increases the complexity of asynchronous GPU programming.
GPU programmers must carefully orchestrate memory movement, tensor-core execution, synchronization, and pipelining to avoid bubbles~\cite{cutlass_blackwell_fmha,thunderkittens2024,zadouri2026flashattention4algorithmkernelpipelining}.
Second, monolithic kernels pack dependency management and overlap decisions into a single opaque schedule and execution unit~\cite{bauer2011cudaDMA,bauer2014Singe}. This creates utilization bubbles and prevents opportunistic execution across kernel boundaries based on runtime resource availability.
Dynamic workloads further exacerbate this limitation, where changing inputs quickly invalidates compile-time overlap and resource-allocation decisions~\cite{soi2025optimalsoftwarepipeliningwarp,hong2023flashdecodingpp}.

Recent frameworks embrace new programming paradigms, libraries, or compilers to ease the programming and performance tuning for asynchronous GPUs.
For example, CUTLASS uses warp specialization to overlap memory and compute within a kernel, while Mirage Persistent Kernel (MPK) ~\cite{cheng2025mpk} and ThunderKittens (TK) ~\cite{thunderkittens2024} extend this idea with cross-task pipelining.
However, these approaches only mitigate, rather than remove, the mismatch between monolithic kernels and asynchronous hardware.
They improve overlap inside kernels or across statically planned tasks, but still preserve the kernel/task as the main unit of composition.

To better match asynchronous GPUs, we propose a direct but fundamental shift in GPU programming and execution model:
Expose each asynchronous hardware unit on GPUs as an isolated programming and execution unit.
Developers therefore program asynchronous units directly and independently, rather than wiring them inside a single monolithic abstraction.
At runtime, each hardware unit is driven by resource-isolated and opportunistic scheduling rather than static orchestration.

Based on this idea, we build \emph{virtual decoupled cores} (\sys{}), the first system providing a decoupled programming interface and execution stack for asynchronous GPUs.
\sys{} realizes the decoupled model by virtualizing memory and compute resources into software-managed execution units: \textit{virtual memory cores} (VMCs) and \textit{virtual compute cores}  (VCCs), each of which could execute independently.
Given operators from high-level ML frameworks, \eg\ PyTorch, \sys{} lowers workloads into dependency-connected \mop{} streams, and maps them onto corresponding virtual units for execution. 

\sys{} provides a unified solution to asynchronous hardware.
First, it lets programmers express computation and data movement as fine-grained and resource-isolated \mop{}s, without manually and statically specifying overlap, orchestration, or synchronization strategies.
Second, at runtime, the system can \textit{automatically overlap} memory and compute operations based on actual dependency satisfaction and resource availability.
Finally, once a library of fine-grained \mop{}s is built for this virtual architecture, future tasks can \emph{reuse and re-orchestrate} them to realize different execution plans, without rebuilding specialized kernels for every workload variant.


Despite its promise, major challenge lies in efficiently realizing the decoupled model on existing GPUs. Modern GPUs still use a SIMT programming interface that favors regular, uniform control flow to expose asynchronous units. Materializing a decoupled runtime on top of this substrate can resemble building a software microarchitectural emulator: it must explicitly track resource availability, manage fine-grained dependencies, and schedule work across asynchronous units in real time.
Such logic is inherently branch-heavy, stateful, and synchronization-intensive, making it a poor fit for SIMT execution.
Without careful design, these control and bookkeeping overheads can easily erase the flexibility and utilization gains of decoupling.
\sys{} addresses this performance challenge with the following key ideas:

First, \sys{} fully exploits the different forms of parallelism available on GPUs to sustain high \mop{} throughput. Rather than interpreting \mops{} through a purely scalar control path, \sys{} organizes each virtual core as a software pipeline with separate control and execution stages, and uses SIMT threads cooperatively within each path for instruction handling, address generation, allocation, and data movement. This design lets \sys{} retain the flexibility of decoupled execution while still issuing and executing \mops{} fast enough to fullfil the capacity of GPU computation and memory.

Second, \sys{} reduces the cost of dependency tracking through co-design of the programming model, \mop{} generation, and execution.
The key insight is to restrict and structure inter-\mop{} dependencies so they can be encoded compactly and partially resolved without bookkeeping.
This allows the \mop{} generator to simplify the dependency handling offline, avoiding much of the expensive cross-virtual-core synchronization at runtime.

We implement \sys{} on asynchronous GPUs with different architectures including NVIDIA GH200, H100, and RTX 6000 Pro GPUs.
We evaluate it with four representative LLM inference workloads. As highlighted in \autoref{fig:perf-intro},
\sys{} achieves \textbf{24\% higher decoding throughput on average} than state-of-the-art hand-optimized kernel and megakernel baselines,
and improves performance further by up to 77\% under dynamic inputs.
At the same time, \sys{} reduces kernel programming and specialization effort by 90\%.
We have open sourced \sys{} at \url{https://github.com/vdcores/vdcores}.

\section{Motivation}
\label{sec:motivation}

\begin{figure*}[!t]
\begin{minipage}{0.48\textwidth}
    \centering
    \includegraphics[width=\columnwidth]{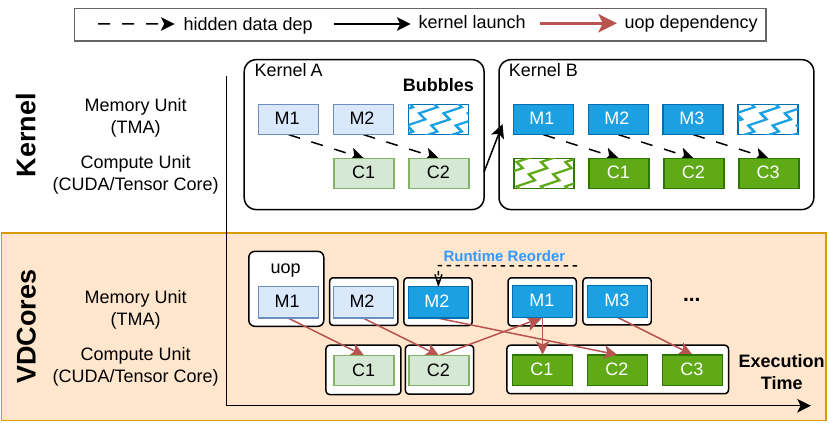}
    \caption{Asynchronous unit utilization under the kernel execution model. \footnotesize{\sys{} overcomes this limitation with asynchronous and independent execution.}}
    \label{fig:motiv}
\end{minipage}
\hfill
\begin{minipage}{0.48\textwidth}
\centering
\includegraphics[width=\columnwidth]{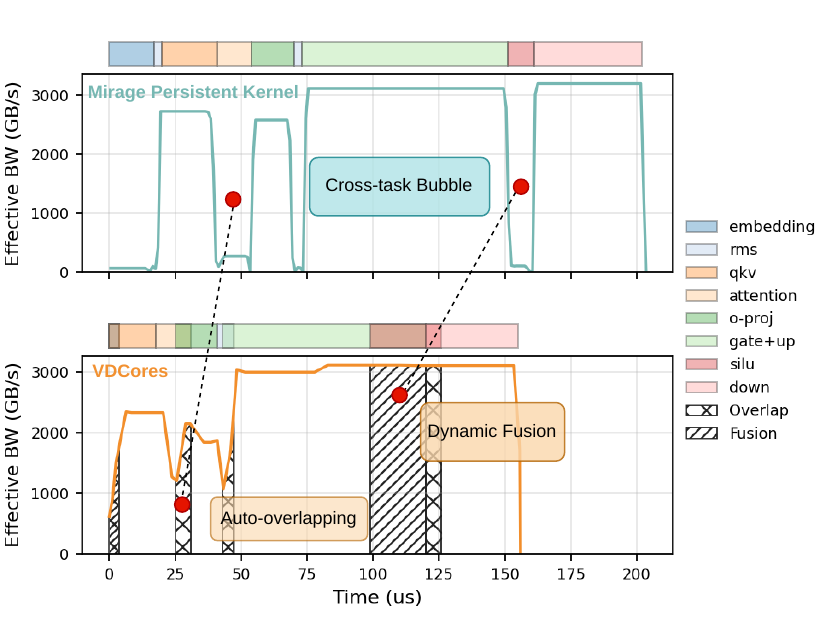}
\caption{Memory-bandwidth utilization comparison on a single \texttt{Llama3-8B} layer. \footnotesize{MPK underutilize memory bandwidth due to inefficient inter-task overlapping.}}
\label{fig:intro-gap}
\end{minipage}
\end{figure*}

\subsection{Asynchronous GPU Programming and Performance}

Existing ML frameworks ~\cite{paszke2019pytorch} lower high-level operators into a sequence of GPU kernels and launch them on the device.
A kernel is the standard GPU programming and execution unit: a host-launched device function that executes the same program across many threads over different data elements.

This abstraction is becoming increasingly strained as GPU hardware grows more heterogeneous and asynchronous.
As shown in Figure~\ref{fig:vdcores-programming-model}, achieving high performance increasingly requires fully utilizing asynchronous units (\eg, Tensor Cores and TMAs) at the same time.
This trend reflects a broader architectural shift in GPUs (\eg, Blackwell~\cite{nvidia_blackwell_2024}, AMD~\cite{amd_cdna_async})
and accelerators (\eg TPU~\cite{jouppi2017tpu}, Groq~\cite{groq_lpu}) toward asynchronous execution.

However,
on asynchronous GPUs, the kernel abstraction creates substantial programming complexity.
The difficulty appears along two fronts.
First, programmers must manually coordinate heterogeneous warp roles inside SIMT code, \eg, using warp specialization to embed producer and consumer behaviors into conditional control flow and explicitly synchronizing their overlap.
Second, performance depends on manually orchestrating deep software pipelines, \eg, multibuffering.
Existing solutions reduce parts of this burden but do not eliminate the require of manual, static and monolith orchestration. 

Further, the monolith kernel abstraction packs asynchronous resources into a single opaque execution unit, itself limits the utilization of asynchronous units.
A kernel bundles multiple resources and execution phases under a single start and finish, so the hardware cannot reuse partially idle resources across kernel boundaries for sub-operations, causing bubbles in execution.
In Figure~\ref{fig:motiv}, between two adjacent kernels mapped to the same hardware unit, one kernel may still be performance writing back while some of its resources, such as the compute engine, are already idle, and cannot be used by the next kernel.
Existing solutions like programmable launch control and prologue/body/epilogue staging only mitigates this overhead.
As in Figure~\ref{fig:intro-gap}, even MPK applies inter-task prologue-body overlapping, significant task-boundary-aligned drops in memory utilization are still visible during execution, indicating that much of the available hardware parallelism remains difficult to express and easy to leave underutilized.


\subsection{Kernel Specialization Efforts}

Modern ML execution is no longer a fixed sequence of uniform operators over stable input shapes.
\eg, S-LoRA~\cite{sheng2024sloraservingthousandsconcurrent} shows that serving many adapters introduces dynamic weights, varying ranks, and heterogeneous batches, requiring both flexible memory management and specialized execution strategies.



Monolithic kernels struggle to adapt to dynamic cases:
a kernel optimized for one regime (\eg, low-latency decoding or single-adapter execution) often performs poorly in another (\eg, long-context decoding, large heterogeneous batches, or mixed-adapter serving).
As a result, efficient execution increasingly relies heavily on \emph{kernel specialization}, including kernel fusion, and megakernel designs.

However, much of this specialization on asynchronous GPUs changes only how resources are orchestrated, including fixed operator bundles, buffer pipeline adjusting, and nearby producer--consumer fusion.
\eg, the fusion support interfaces in Megatron-Core~\cite{shoeybi2019megatron} (11 fusion modules) and vLLM~\cite{kwon2023pagedattention} (9 fusion passes) are dominated by epilogue/prologue-style fusion.
This specialization is simple in logic, but creates a huge programming overhead with a monolithic kernel model, which creates a cross-product problem: even small orchestration changes can cause large performance shifts and therefore require new kernel variants.

\boldpara{Beyond the monolithic kernel abstraction.}
Taken together, these limitations call for a direct, resource-facing model for asynchronous GPUs. Rather than packaging resources, the model should expose them as independent work units that can be composed explicitly and scheduled opportunistically at runtime.

\begin{figure}[t]
\centering
\includegraphics[width=0.8\columnwidth]{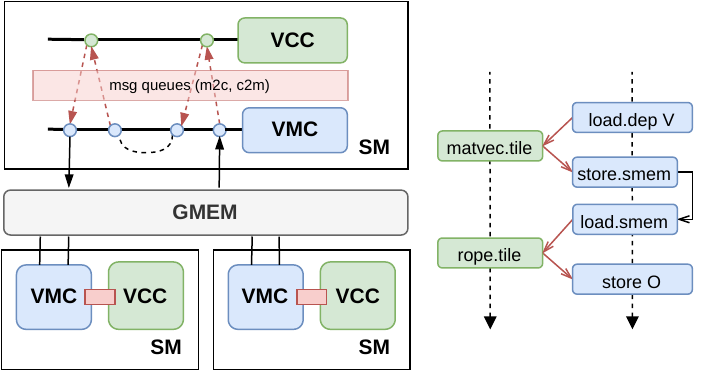}
\\[-0.2em]
{\caption{\sys{} Abstract machine model.}\label{fig:system-overview}}
\end{figure}

\section{\sys{} Overview}

\sys{} serves a role similar to CUDA in GPU programming:
it provides a programming abstraction and execution model for asynchronous GPUs (\S~\ref{sec:design:model}).
It is built around a new hardware execution-unit abstraction, \textbf{decoupled cores}, including \textit{virtual memory cores} (VMCs) and \textit{virtual compute cores}  (VCCs), and a new programming and execution unit, \textbf{micro-operators (\mops{})}.

\sys{} serves as a drop-in replacement for the CUDA backend.
For GPU applications, \eg, performing matrix multiplication in PyTorch,
CUDA converts them into a list of kernel launches,
while \sys{} converts them into a \textit{\mop{} graph} (\S~\ref{sec:design-sched}), consisting
of \mop{} streams for virtual cores,
defining the detailed operations to execute, 
and
connected by \textit{dependency edges}, defining the execution order of the \mops{},
as illustrated in \autoref{fig:system-overview}.
\sys{} then submits the constructed \mops{} to its on-GPU executors, \textit{virtual} decoupled cores (\S~\ref{sec:design:executor}), for execution.
Within each virtual core, \mops{} execute in a dependency-driven manner, following dataflow order~\cite{arvind2003dataflow} rather than instruction order.
After completing one \mop{}, the executor selects another ready \mop{} for execution and continues until the entire \mop{} graph finishes.

\begin{figure}[t]
\centering
\includegraphics[width=\columnwidth]{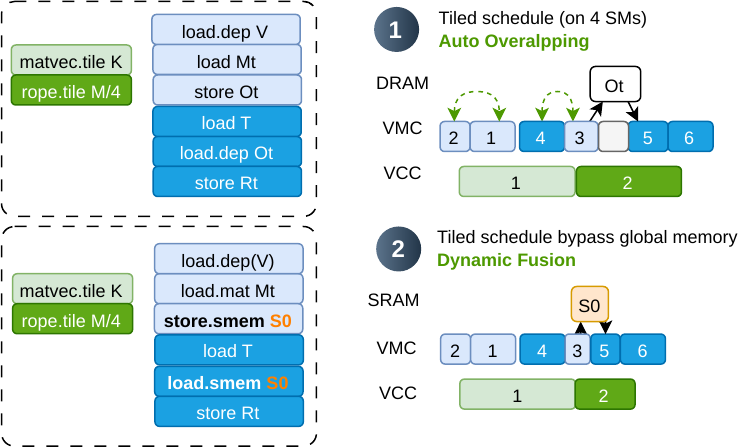}
\caption{\sys{} Execution Example.}
\label{fig:overview-example}
\end{figure}

\boldunderpara{Execution Example.}~~~
We illustrate \sys{} execution flow using an instruction flow of two ML operators: a matrix-vector multiplication ($O = M @ N$), connected by a RoPE rotation ($R = rope(O,T)$), as shown in Figure~\ref{fig:overview-example}.

The \sys{} scheduler decomposes operators into a a list of dependency-connecting compute and memory \mop{}s.
With 4 pair of VCCs and VMCs available, \sys{} decomposition is to tile the matrix along the $M$ dimension and assign the work to each virtual core pair.
Each virtual core's \mop{} sequence is shown as \blackcircled{1}.
On each virtual core, \sys{} execute \mop{} based on dependency and resource readiness.
We illustrate the process and it's benefits with two execution examples.

First,
\sys{} virtual cores reorders the execution to fill execution idle gaps,
\ie, \textit{auto-overlapping} between \mop{}s.
In \blackcircled{1}, the \texttt{load Mt} \mop{} can be reordered and executed before \texttt{load.dep V} because the former have no dependency, while \texttt{load.dep V} is could still be waiting on its dependent instructions to finish.
On \sys{}, any \mop{} with ready dependency be executed first, reducing the runtime stalls casued by static scheduling and coarse-grained dependencies.

Further, \sys{} \mop{} builder (\S~\ref{sec:design-sched}) exploits this dynamic execution model to generate \mop{} sequences with fewer dependency-caused stalls,
enabling optimization across operator boundaries.
One example is \textit{dynamic fusion},
which promotes global-memory communication to shared memory for one-to-one dependency edges between memory \mop{}s on the same virtual memory core.
As shown in schedule \blackcircled{2},
the \sys{} \mop{} optimizer detects this pattern and rewrites \texttt{store} and \texttt{load} into \texttt{store.local} and \texttt{load.local}.
At runtime, the dependency is resolved through local queues with low overhead.
In effect, this achieves operator fusion without requiring a manually written fused variant.

\section{\sys{} Design}
\label{sec:design}

In this section, we introduce the three major components materialize the decoupled programming and execution model on GPUs: \sys{} decoupled model, \sys{} \mop{} generator and \sys{} virtual cores.

\subsection{The \sys{} Decoupled Model}
\label{sec:design:model}

This section introduces the decoupled programming model \sys{} adopt and how it shifts the programming and execution paradigm on asynchronous GPUs.

\begin{figure}[t]
\centering
\begin{minipage}{0.96\columnwidth}
\begin{lstlisting}[language=C++,
  emph={[1]m2c,c2m,alloc_registers},
  emphstyle={[1]\color{blue}\bfseries},
  emph={[2]ctx,inst},
  emphstyle={[2]\bfseries},
  emph={[3]push,pop_wait},
  emphstyle={[3]\color{magenta}\bfseries},
  basicstyle=\scriptsize\ttfamily,
  frame=tb,
  numbers=left,
  numberstyle=\scriptsize,
  stepnumber=1,
  numbersep=6pt]
VDC_DEFINE_UOP(COP_MATVEC, h_matvec)
__vdc_cop__ h_matvec(VdcInst &inst, VdcCtx &ctx) {
  auto c = ctx.alloc_registers(NCREGS);
  // the number of tiles are encoded in size field
  for (int i = 0; i < inst.size; i++) {
    float *a = ctx.m2c.pop_wait();
    float *b = ctx.m2c.pop_wait();
    matvec_tile_accumulate(a, b, c);
    ctx.c2m.push(a, b);
  }
  float *c_shared = ctx.m2c.pop_wait();
  copy(c, c_shared);
  ctx.m2c.push(c_shared);
}
\end{lstlisting}
\end{minipage}
\caption{Simplified new compute \mop{} \texttt{matvec}. \footnotesize{Kernel programmers use dependency queues \texttt{m2c} and \texttt{c2m}, to acquire and release memory resources with \texttt{push}/\texttt{pop}.}}
\label{fig:design-add-new-mop}
\end{figure}

\subsubsection{\sys{} Abstraction of GPU Hardware}
\label{sec:model:abstract}

\sys{} decomposes GPU hardware execution units (streaming processors, SMs, or compute units, CUs) into resource-specialized cores: Compute Cores and Memory Cores.
Each core resembles a lightweight register machine; \sys{} further specifies its architectural state (\eg, register file, local-memory management) and programming interface to access them.
Each type of core executes a flow of specific micro-operations (\mops{}): Compute Cores execute compute \mops{}, while Memory Cores execute memory \mops{}.

A \mop{} is the smallest unit of execution and programming unit in \sys{}.
Each \mop{} describes a unit of task within the capability of the execution unit when the resource is ready.
Memory \mops{} in \sys{} include memory movement and management operations, \eg, \texttt{load} and \texttt{store}.
For example, the memory \mop{}, \texttt{load.dep} in \autoref{fig:system-overview} loads a memory block from \texttt{addr} in global memory to the local memory.
Compute \mops{} are computation atoms that can be completed within a single resource domain (\eg, a local tensor core) and encode tile-granular semantics.
For example, the compute \mop{} \texttt{attn} performs the fused attention score computation.
Control \mops{} implement control flow like conditions and branching.
\sys{} further provides control \mops{}, including \texttt{loop}, \texttt{continue\_if}, which can execute on all virtual cores.
Together, these \mops{} give each virtual core a small programmable execution context, enabling it to express complex GPU computation patterns beyond a fixed instruction stream.

\subsubsection{\sys{} Programming Model}

Similar to programming kernels in CUDA, on \sys{}, GPU programmer programs \uop{}s to extend \sys{}'s capabilities or support new machine learning operators,
\eg, a new attention operator such as \texttt{LinearAttention} or new memory operations for data movement over GPU-GPU interconnects.

As shown in \autoref{fig:design-add-new-mop}, adding a new \mop{} involves two steps.
First, the developer defines a new opcode and specifies how the fields in the \mop{} instruction word should be interpreted.
Second, the developer implements and registers a GPU device-side \mop{} handler that executes the \mop{}. It takes the instruction word and the virtual core context as input.
Programmers uses the context to request the service of runtime, including resolving dependency, allocating resource and changing register states.
As shown in \autoref{fig:design-add-new-mop}, it can use the \texttt{m2c} and \texttt{c2m} dependency queues to acquire memory resources from memory cores and release them after use.
Compared with writing monolithic kernels, adding a new \mop{} frees the programmer from manually orchestrating asynchronous coordination across hardware resources.

\sys{} pre-defines a list of compute, memory, and shared control \mop{}s that support common types of ML operations.


\subsubsection{\sys{} Execution Model}

\sys{} implies dependency-driven execution model.
Each \mop{} may depend on resources produced or owned by other virtual cores in order to complete its task.
For example (\autoref{fig:design-add-new-mop}), a \texttt{matvec} \mop{} requires shared-memory tiles containing the input matrix and vector tiles during execution.
In \sys{}, such relationships are represented explicitly as dependencies edges between \mop{}s, marking a execute-after relationship.


This execution model improves asynchronous hardware utilization in two ways.
First, by removing ML-operator execution boundaries and merging work into a single \mop{} graph, it prevents resources from being blocked behind unrelated work.
Second, by resolving execution at runtime from actual resource readiness, it naturally adapts to dynamic execution conditions instead of relying on a fixed static schedule.

However,
\sys{}'s dependency-driven execution model can be expensive to realize on GPUs,
due to the major overhead of dependency tracking.
Conventional dataflow and out-of-order architectures \cite{arvind1986dataflow} rely on dependency scoreboards that track every issued operation, incurring substantial cycle and memory overhead on GPUs.
\sys{} addresses this challenge in two ways.
First, it reduces the complexity of the dependency graph by restricting and simplifying the dependency model.
Second, instead of relying on runtime, cross-core dependency resolution, \sys{} uses co-design to encode dependency structure into \mop{}s ahead of time and to localize dependency resolution as much as possible, while preserving the same execution flexibility.

\boldpara{Encoding dependency in \mop{}.}
\sys{} restricts each memory \mop{} to have at most one inter-memory-\mop{} dependency, plus an optional dependency to computation \mop{} executing on the same SM.
This simplification eliminates the need to represent and resolve an arbitrary dependency graph in hardware.
Instead, \sys{} encodes dependency information directly into \mop{}s in a GPU-friendly form.

Dependencies between \mop{}s in local memory core and compute cores are represented using \texttt{flags} on the memory \mop{} side. For instance, when a memory \mop{} carries a \texttt{send} flag, it forwards its allocated memory region to a local compute core.
Conversely, a \texttt{recv} flag indicates that the \mop{} consumes a region sent from a local compute core.

Similarly, dependency between two \mops{} are also encoded into a \texttt{depId} field, and its direction is determined by the \texttt{send}/\texttt{recv} flags.
Thus, when generating \mop{}s, \sys{} does not store explicit pointers to dependent \mop{}s, and instead tracks dependencies entirely through these encoded fields.

\boldpara{Virtual-flow-based dependency-driven \mop{} execution.}
To enables the runtime to quickly identify \mop{}s that are independent and can therefore execute in parallel, without performing general dependency checks at runtime.

To support this, \sys{} \mop{} generate (\S~\ref{sec:design-sched}) assign each \sys{} \mop{} a \texttt{virtualFlowId} that encodes coarse-grained dependency structure among \mop{}s.
\mop{}s connected by direct dependency edges are assigned to the same virtual flow, while independent \mop{}s are placed in different virtual flows.
At runtime, instructions within the same virtual flow execute in order, whereas instructions from different virtual flows may be reordered and overlapped when resources permit.

\subsection{\sys{} \mop{} Generator}
\label{sec:design-sched}

\sys{} \mop{} generator takes highlevel machine learning operator graph (\eg, PyTorch computation DAG) and convert them into \mop{}, assign them to virtual cores and build dependency between them.
\sys{} resolves this progressively across two tiers to balance scheduling effectiveness and scheduling overhead.
First, it chooses how to decompose ML operators to jobs on multiple cores.
With a global view of the cross-operator DAG, this level focuses on exposing task-level parallelism and balancing load across virtual cores.
Then, it optimizes the generated per-core \mop{} flows.
This level focuses on memory-\mop{} placement, fine-grained dependence resolution, and instruction execution efficiency within each virtual core.


\begin{figure}[!t]
\centering
\centering
\includegraphics[width=\linewidth]{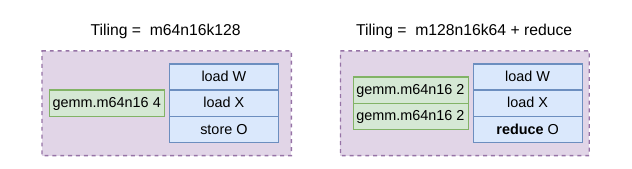}
\captionof{figure}{\sys{} Composes \mop{}s to Dynamically Tiles Shapes.}
\label{fig:dynamic-tile}
\end{figure}

\subsubsection{Adaptive Scheduling and Tiling}



\sys{} first schedules the DAG of logical operators to virtual cores, similar to existing GPU backends.
The key difference is that \sys{} schedules over the full DAG (instead of single DAG node) and adapts execution at runtime by composing \mop{}s directly, rather than by selecting from precompiled specialized kernels.
We refer to this flexibility as \textit{adaptive scheduling}. 
Examples include partitioning a GEMM along the $M$, $N$, or $K$ dimensions, or partitioning attention across token and head blocks.
Different tilings results on different workload to be assigned on each computation unit,
where \sys{} could dynamically build through composing \mops{} instead of requiring a new kernel. 

Given this flexibility, \sys{} chooses tilings by iteratively eliminating dominant critical paths in the execution DAG.
Each operator is associated with a set of valid decompositions into \mop{}s, provided by the \mop{} builder, and each \mop{} and dependency edge carries a cost model that estimates its execution cost on virtual cores.
A critical path is considered dominant if its cost exceeds the average per-core cost by more than a threshold.
The scheduler then refines operators on these paths by changing their tilings until the expected benefit becomes marginal or no dominant critical path remains.
This process drives globally optimized execution while also reducing direct dependency-resolution overhead.

\subsubsection{Dependency Graph Lowering} 

\sys{} next embeds inter-\mop{} dependencies directly into the per-virtual-core \mop{} bytecode. 
It first performs virtual flow assignment, decomposing the dependency graph on each virtual core into independent chains and assigning one virtual flow to each chain. 
This decomposition remains possible even with control flow loops and branches, because computation \mop{}s introduce no direct dependency between each other, and each memory \mop{} has at most one incoming inter-memory dependence.
As a result, independent \mop{}s can bypass stalled ones instead of being serialized behind them in a single producer pipeline.

\sys{} then checks for and eliminates deadlocks by elaborating resource allocation along \mop{} flow on each core, especially slot-based local memory allocation, and detecting potential over-allocation.
When necessary, it reorders instructions without violating dependence constraints.
For example, it may move an independent store ahead of a later load to free resources earlier.
This step is particularly important when one VMC serves multiple VCCs.

Finally, once dependencies are embedded, \sys{} removes redundant dependency checking. When two \mop{}s are placed on the same virtual flow, in-order execution already enforces their readiness order, so explicit barriers implied by transitive dependences can be eliminated. This reduces polling and control overhead while preserving correctness.

\subsubsection{Dynamic Data Placement and Fusion}
Several optimizations of \sys{} \mop{} generation requires multiple steps to work together, including dynamic data placement.
dynamic data placement change the  placement of memory \mop{}s between neighboring \mop{}s with one-to-one dependency. For example, intermediate data can be promoted to shared memory to enable fusion-like locality, or kept separate to expose more overlap opportunities.
The scheduler is aware of this optimization and accounts for the reduced cost of one-to-one dependencies.

This design differs from conventional kernel fusion. Because \sys{} already decouples memory movement and computation into separate \mop{}s, composing them requires neither code generation nor recompilation. The same computation can therefore be realized either as a locality-oriented fused pipeline or as a more weakly coupled flow that exposes additional overlap opportunities.

\subsection{\sys{} GPU Executor}
\label{sec:design:executor}

\begin{figure}[!t]
\centering
\centering
\includegraphics[width=0.88\linewidth]{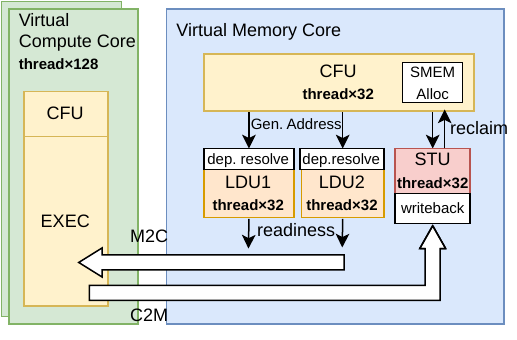}
\captionof{figure}{\sys{} virtual executor. \footnotesize{The example shows executors on a single H100 SM, with two VCC and single VMC.}}
\label{fig:vdcore-full-execution-flow}
\end{figure}



\label{sec:design-runtime}

Virtual cores are software \mop{} interpreter built on top of GPU hardware execution units.
Each virtual core is launched once as a persistent kernel at the beginning of execution.
\sys{} runtime streams new \mop{}s to virtual cores for execution when new requests arrive.

A core design challenge of virtual cores is \textit{how to preserve the flexibility of \sys{} decoupled model without introducing excessive runtime overhead}.
\sys{} addresses this challenge by drawing inspiration from classical microarchitecture while restructuring the design to match GPU hardware.
At a high level, it forms a software analogous to a pipelined superscalar microarchitecture: multiple virtual execution units are organized as a wide pipeline,
with each unit mapped onto GPU warps and register state,
while explicitly leveraging SIMT parallelism within them.

\subsubsection{Virtual Memory Core}
\label{sec:design-runtime-memory}
Virtual memory cores (VMCs) execute memory and control \mop{}s and manage their internal state (\eg, loop counters) and memory resources (\eg, shared-memory capacity within an SM). 
VMC \mop{} execution throughput is critical to \sys{} performance.
For example, an NVIDIA H200 GPU has up to 4TB/s DRAM bandwidth.
The \sys{} runtime must sustain over 10K tile-level load/store \mop{}/s  to avoid becoming the performance bottleneck.


VMC employs two levels of parallelism.
The first comes from building an 2-stage execution pipeline for each \mop{}.
Each \mop{} is first processed by a front-end control-flow unit (CFU), which manages register state, handles control \mops{}, and performs address generation.
The decoded \mop{} and generated address is then issued to multiple parallel load units (LDUs) and store units (STUs),
where dependency resolution and memory movement are handled independently.
This pipelined design reduces the initiation interval for processing a \mop{}: throughput is determined by the slowest pipeline stage rather than the sum of all stages, and instruction sequences no longer suffer from head-of-line blocking.
Compared with placing all work in a single GPU loop, this design improves effective \mop{} throughput by $4.2\times$.

Within each pipeline stage, VMC further parallelize control and data movement by leveraging parallel GPU threads.
In the CFU, thread-level parallelism and software pipelining allow \mop{} decode and address generation to proceed concurrently with instruction fetch and register access.
Managing local memory is likewise parallelized through a SIMT allocator.
Each \mop{} may request an arbitrary number of contiguous slots, and allocated slots may be freed in an order different from allocation (\eg, when implementing FlashAttention \mop{}).
VMC tracks memory availability as a 32-bit bitmask in shared memory and uses multiple threads to search candidate allocation positions in parallel.
Each allocator thread probes for a slot starting from its thread ID, and the warp then uses a warp-level voting primitive to identify the first available location.


\boldpara{Parallel virtual flow execution}: 
by dynamically mapping virtual flows onto physical execution units.
At runtime, the CFU maintains a mapping from each virtual flow to a physical execution unit.
For a given \mop{}, the CFU dispatches it according to this mapping.
\mops{} within the same flow are always assigned to the same execution unit and execute in order, preserving intra-flow ordering.
When the number of active virtual flows exceeds the number of physical units, the CFU multiplexes multiple flows onto the same unit. This design preserves dependency semantics independent of the underlying runtime implementation.

Non-dependent operations can execute in parallel across multiple LDUs and STUs.
However, within each individual LDU or STU, \mops{} execute strictly in order, following the sequence of dependency resolution, operation execution, and dependency message delivery.
As a result, a \mop{} may spin-wait on an unresolved dependency and temporarily block later \mop{}s assigned to the same unit. For example, within an LDU, one \mop{} may wait on a pending dependency and thereby stall subsequent \mop{}s in that unit.
This design preserves ordering and dependencies without creating a single hot spot of dependency tracking in VMCs.


\subsubsection{Virtual Compute Core}
Virtual compute cores handle compute resources, including register files, SIMT cores and asynchronous matrix-computation units (\eg, Tensor Cores).
Each VCC contains a control-flow unit (CFU) and multiple execution units (EU) for executing computation \mop{}s. 
As an example, in \autoref{fig:vdcore-full-execution-flow},
\sys{} by default create two EU per SM, each handling 128 SIMT threads and attached CUDA core resources, when applicable.


Unlike VMC, the CFU and EU in VCC are not mapped to separate parallel threads.
Instead, they are organized as a software pipeline.
The CFU state is stored in EU-shared memory, and for each instruction,
the CFU runs first on EU's thread set, maintaining and preparing state (\eg, updating the GPR value), in a manner similar to an operating-system kernel, before yielding to the EU to execute the target \mop{}.
This design is motivated by the low density of control \mop{}s in VCC execution: more than 98.6\% of execution time is spent in the EU performing actual \mop{} execution.

\subsubsection{Dependency Resolving}

On \sys{}, dependency are resolved at runtime by passing dependency messages through first-in first-out (FIFO) messages queues connecting virtual cores.
Following \sys{} execution model, VCCs only communicate with VMCs on the same execution unit, while all VMCs can talk to each other.

VMC-to-VCC messages include shared memory regions that have been loaded from global memory and thus can be used by the compute units.
VCC-to-VMC message queues return used or generated memory regions to be recycled or written back.
Similarly, VMCs communicate through a global queue allocated based on its \texttt{depid}. Between VMCs, messages passes the ownership of global memory resources (\eg, after writing to a global memory block).

\subsubsection{Avoiding Deadlock.}

Dependency-driven execution introduces a risk of deadlock when outstanding \mop{}s across virtual cores form a cyclic wait relationship on shared resources. A typical case arises between a VMC and a VCC: the compute side waits for inputs to be loaded into shared memory before producing outputs, while the memory side waits for memory slots to be released before issuing additional \texttt{load} \mops{}.

Since \sys{} forbids cyclic \mop{} dependencies, deadlocks can only arise from resource-allocation cycles rather than dependency cycles.
\sys{} prevents such deadlocks by linearizing resource acquisition through a joint scheduling and runtime protocol, summarized as \textit{in-order allocation and out-of-order execution}.
The \sys{} \mop{} generator (\S~\ref{sec:design-sched}) first constructs a baseline \mop{} order that is deadlock-free under in-order execution. The runtime then preserves this order during local resource allocation (\eg, shared-memory slots), while still allowing later execution steps to proceed out of order once their dependencies are satisfied (\eg, loading a memory slot and sending notification).
This avoids deadlock because resource allocation remains acyclic in the preserved baseline order, and once allocated, blocked \mops{} wait only on acyclic dependencies, so some ready \mop{} can always execute and eventually release resources.

\section{Implementation}

\begin{figure*}[!t]
\centering
\includegraphics[width=\textwidth]{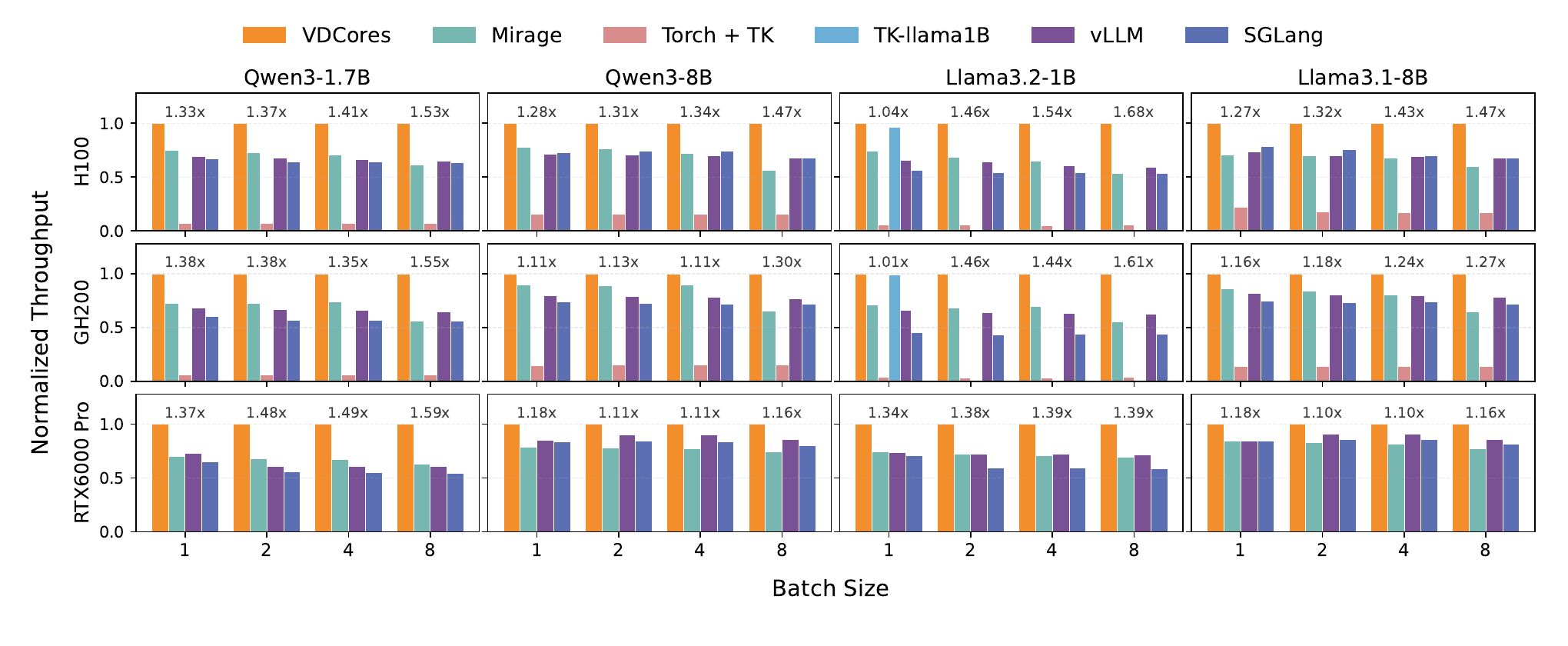}
\vspace{-0.5in}
\caption{End-to-end decoding performance over LLM inference. \footnotesize{Throughput is normalized to \sys{}. The number on each group shows \sys{}'s performance gain over the best baseline system. TK: ThunderKittens.}}
\label{fig:eval-end2end-matrix}
\end{figure*}

We implement \sys{}'s executors and \mop{} generator for NVIDIA Hopper and Blackwell GPUs in 5K lines of C++/CUDA.
On top of this runtime, \sys{} provides both low-level \mop{} execution APIs and high-level ML-operator execution interfaces in 4K Python LoC, enabling integration with PyTorch as a backend.

\begin{table}[t]
\centering
\small
\setlength{\tabcolsep}{4.0pt}
\begin{tabular}{lcccc}
\hline
\textbf{GPU} & \textbf{\#SMs} & \textbf{ShMem KB/SM} & \textbf{DRAM BW}  \\
\hline
\hline
H100 & 132 & 228 KB & 3.35 TB/s \\
GH200 & 132 & 228 KB & 4.00 TB/s \\
RTX6000 Pro & 188 & 100 KB & 960 GB/s \\
\hline
\end{tabular}
\vspace{0.3em}
\caption{Hardware configuration used in end-to-end evaluation. On each system \sys{} deploys 1 VMC and 2 VCCs per SM, with 8KB memory slot size.}
\label{tab:eval-hardware-spec}
\end{table}

We highlight the following implementation details further help \sys{} to improve execution efficiency:

\boldpara{Portable Runtime.}
We port the runtime on three different NVIDIA GPUs with different microarchecture specifications, showing the portability of \sys{}.
As summarized in \autoref{tab:eval-hardware-spec}, each of them has different number of SMs, shared memory size, and global memory bandwidth.
They also feature different asynchronous execution unit configurations. \eg, GH200 and H100 additionally support asynchronous tensor core execution with \texttt{WGMMA} instructions.

These difference is hidden by the \sys{} virtual core and \mop{} abstraction.
same \mop{} graph executes on both architectures, while \texttt{matvec} \mop{} are backed by different implementations.

\boldpara{Message Passing on Virtual Cores.}
Both dependency queues and inter-execution unit communication in \sys{} are implemented through message passing.
We realize this mechanism using $mbarrier$ instructions ~\cite{cuda_async_barriers}, which provide efficient synchronization while enabling space-sharing of hardware resources.
Inside each SM, all execution units, including VMC and VCCs, share the same four physical execution slots. If an execution unit blocks on a message (\eg, at \texttt{pop\_wait()}), it is de-scheduled and resumed only when the awaited message arrives. As a result, \sys{} avoids spin waiting and reduces wasted cycles from idle execution units.

\boldpara{Dynamic address generation}
\sys{} supports dynamic address generation for every \mop{} to produce runtime-dependent addresses.
In \sys{}, an instruction may be marked with a \texttt{dynamic} flag, indicating that its \texttt{address} field (which may encode an address, an $n$-dimensional tensor coordinate, or the index of a TMA descriptor)
is generated dynamically and needs to be computed by adding an offset from an accumulator register.
Further,
the accumulators are designed to be updated by control \mop{}s, \eg, by \texttt{repeat} based on loop counter.
This mechanism substantially reduces the number of instructions needed for recurring patterns, which are common in LLM inference workloads,
\eg, full inference of \texttt{llama-8b} is encoded in only 224 \mop{}s per virtual core.



\section{Evaluation}
\label{sec:results}
We answer three key questions in this section:
\begin{enumerate}[leftmargin=*]
    \item Does \sys{} improve end-to-end model execution over state-of-the-art kernel and megakernel systems? (\S~\ref{sec:eval-end2end})
    \item How much does each key optimization in \sys{} contribute to the performance gain? (\S~\ref{sec:eval-perf-deepdive})
    \item Is software-managed decoupled execution practical given the runtime overhead introduced by \sys{}? (\S~\ref{sec:eval-ablation})
\end{enumerate}

\subsection{End-to-End Performance and Coding Effort}
\label{sec:eval-end2end}

We first evaluate \sys{} on end-to-end LLM inference under offline decoding. We study four representative models—Qwen3-1.7B, Qwen3-8B, Llama3.2-1B, and Llama3.1-8B.
Our setup includes KV cache and paged attention. Without continuous batching, we experiment with offline decoding of a fixed batch for 64 steps from a 128-token context. We report token-generation throughput for batch sizes 1 to 8.

We compare against strong baselines spanning the main current design points: vLLM~\cite{kwon2023pagedattention} and SGLang~\cite{zheng2024sglang} represent optimized kernel-per-operator execution with JIT generation and CUDA Graphs, while Mirage~\cite{wu2025mirage} and ThunderKittens-llama1B~\cite{spector2025nobubbles} implementations represent expert-tuned megakernels. We also include Torch+ThunderKittens as a lower-engineering baseline using PyTorch’s execution stack plugged with an optimized attention kernel.

\begin{table*}[t]
\centering
\small
\setlength{\tabcolsep}{3.8pt}
\begin{tabular}{lcccc|c|c}
\hline
\textbf{System} & \textbf{\# Independent} & \textbf{LoC} & \textbf{\# Fused} & \textbf{Fused LoC} & \textbf{Avg. H100 Perf.} & \textbf{Reusability} \\
 & \textbf{kernels/tasks/\mop{}s} &  & \textbf{tasks} &  & \textbf{(norm.)} & \textbf{(for new model and hardware)} \\
\hline
\textbf{\sys} & \textbf{6} & \textbf{741} & \textbf{0} & \textbf{--} & \textbf{1.00} & \textbf{High} \\
Mirage & 8 & 2,339 & 3 & 1,073 & 0.83 & Low (Monolith Tasks) \\
TK-llama1b & 7 & 2,065 & 5 & 596 & 0.95 & Low (BS=1, Special Fusion) \\
vLLM & 14 & 6,424 & 8 & 3,788 & 0.63 & High \\
\hline
\end{tabular}
\vspace{0.2em}
\caption{\label{tab:eval-simplicity-programming-effort}
GPU Program Line of Code Comparison across Systems for implementing \texttt{llama3-1b}, along with qualitative reusability under new inputs and new architectures.}
\end{table*}

\autoref{fig:eval-end2end-matrix} shows the end-to-end throughput results. Across all 48 evaluated combinations of model, hardware, and batch size, \sys{} consistently achieves the lowest per-token decoding latency among both kernel-per-operator and megakernel baselines. Relative to the best baseline in each setting, \sys{} delivers a 1.31x geometric-mean speedup, corresponding to a 23\% average reduction in per-token latency, and up to a 1.68x performance improvement.

The gains are especially pronounced on smaller models, where shorter task durations amplify the cost of underutilization across kernel boundaries. MPK reduces kernel launch overhead through persistent-style execution, but its pipeline structure is still packed into each individual task and thus exposes only limited overlap opportunities. ThunderKittens-Llama1B removes most stalls through manually designed fusion and prefetching specialized for this model architecture and batch-size-one regime, reaching roughly 96\% of \sys{}'s performance. The remaining gap is consistent with our argument that compile-time task-coupled scheduling, even when highly optimized, still leaves some overlap opportunities uncovered.

\boldpara{Programming effort.}
\label{sec:eval-simplicity-programming-effort}
Beyond performance, \sys{} substantially reduces the need for manual specialization. 
As shown in \autoref{tab:eval-simplicity-programming-effort}, \sys{} implements the evaluated Llama3-1B pipeline with only 6 reusable \mop{}s and 741 lines of GPU code, while requiring no task-specific fused implementation. In contrast, Mirage and TK-Megakernel rely on 8 and 7 independent operators, respectively, together with 3 and 5 fused tasks, and require roughly 3x more GPU code. 
This gap reflects a fundamental difference in where optimization complexity lives. Kernel-centric systems realize performance through task-specific fusion and specialized implementations, so both code size and maintenance cost grow as operators, surrounding stages, or target hardware change. By contrast, \sys{} reuses the same small set of \mop{} building blocks and changes only their composition and scheduling, shifting optimization effort from per-task kernel variants to a reusable runtime substrate.

\begin{figure}[!t]
\centering
\centering
\includegraphics[width=\linewidth]{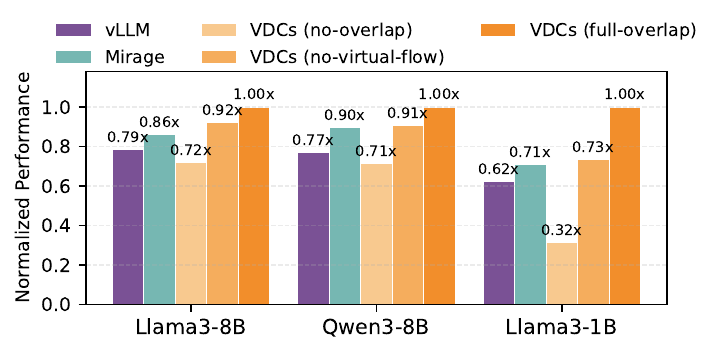}
\captionof{figure}{Auto overlapping deep dive. \textit{VDCs}: Virtual Decoupled Cores.
}
\label{fig:eval-deepdive-core-overlap}
\end{figure}

\subsection{Performance Deep Dive}
\label{sec:eval-perf-deepdive}
To understand where \sys{}'s benefits come from, we evaluate each key mechanism with targeted case studies to examine when each is effective.

\subsubsection{Auto Overlapping}
\label{sec:eval-deepdive-overlap}


\sys{} derives intra- and inter-task overlap directly from dependency-driven execution of \mop{}s. To isolate the benefit of this mechanism, we start from a restricted version of \sys{} that converts \mop{} data dependencies into task issue barriers, forcing all instructions from a later task to wait even when their dependencies are already satisfied. This resembles a kernel-per-operator execution style. We then enable cross-task overlap by restoring fine-grained dependency-driven instruction issue. Finally, we enable virtual-flow assignment, which moderates head-of-line blocking by allowing ready \mop{}s to bypass stalled ones.

As \autoref{fig:eval-deepdive-core-overlap} shows, enabling cross-task overlap accounts for the dominant share of \sys{}’s benefit across all workloads. Without it, \sys{} becomes comparable to, and in some cases slower than, baseline systems, showing that \sys{}’s advantage does not come from faster individual operators but from the overlap exposed by dependency-driven scheduling. Enabling virtual-flow assignment yields a further 5\% improvement on top.


\begin{figure}[t]
\centering
\includegraphics[width=\columnwidth]{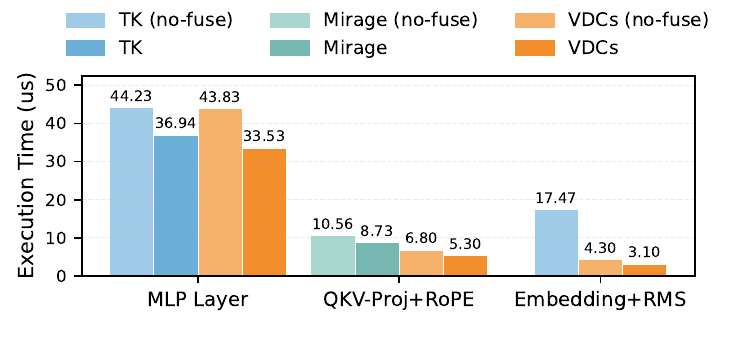}
\caption{\sys{} Dynamic Fusion Compared to Manual Fused Kernels.}
\label{fig:eval-simplicity-dynamic-fusion}
\end{figure}

\subsubsection{Dynamic Fusion}
\label{sec:eval-dynamic-fusion}


Dynamic fusion in \sys{} is realized by rewriting the memory path of intermediate data between neighboring tasks, rather than by introducing a new fused task implementation. We evaluate this mechanism on two representative manually fused patterns, QKV-Projection + RoPE and the MLP block. As \autoref{fig:eval-simplicity-dynamic-fusion} shows, \sys{} closely matches the performance benefit of manual fusion in both cases, demonstrating that composing \mop{}s differently can exploit similar locality benefits without the need for specialized kernels. 

We further study Embedding + RMS, a fusion opportunity that emerges automatically from \sys{}’s runtime placement decisions but is typically left unfused in expert-tuned systems. In this case, fusion reduces execution time from 4.30 to 3.10\textmu s, a 28\% improvement over its own non-fused path

\begin{figure}[t]
\centering
\includegraphics[width=\linewidth]{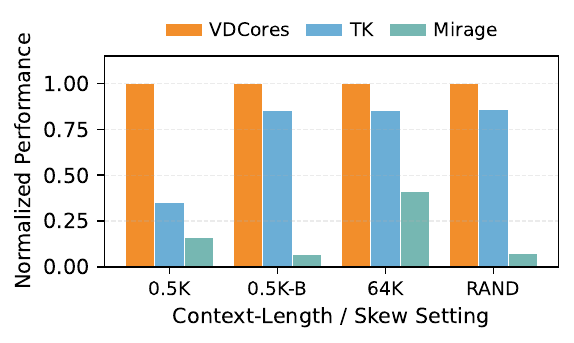}
\caption{Dynamic uneven-context evaluation (H100). \footnotesize{x-axis labels: \texttt{0.5K} = \texttt{1\(\times\)512}, \texttt{0.5K-B} = \texttt{64\(\times\)512}, \texttt{64K} = \texttt{1\(\times\)65536}, and \texttt{RAND} = \texttt{64\(\times\)random[256,1024]}.}}
\label{fig:eval-end2end-uneven-context}
\end{figure}
\begin{figure}[!t]
\centering
\begin{minipage}[t]{0.64\columnwidth}
\vspace{0pt}
\centering
\includegraphics[width=\linewidth]{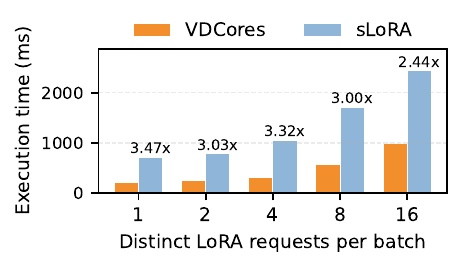}
\\[-0.3em]
{\footnotesize (a) Uneven-context performance}
\end{minipage}\hfill
\begin{minipage}[t]{0.34\columnwidth}
\vspace{0pt}
\centering
\scriptsize
\setlength{\tabcolsep}{2.2pt}
\renewcommand{\arraystretch}{1.03}
\begin{tabular}[t]{@{}l r@{}}
\toprule
System & Sched. Time (ms) \\
\midrule
Mirage & 27000 \\
TK (\textbf{attn}) & 2.28 \\
vLLM & 352 \\
SGLang & 261 \\
\midrule
\sys{} & 1.4 \\
\bottomrule
\end{tabular}
\\
\vspace{0.2in}
{\footnotesize (b) Scheduling time}
\end{minipage}
\caption{Adaptive Scheduling with dynamic LoRA serving. In-batch LoRA-adapter distribution follows.}
\label{fig:eval-runtime-composability-throughput}
\end{figure}

\subsubsection{Adaptive Scheduling}
\label{sec:eval-deepdive-dynamicsched}

\sys{} adapts to changing workloads by recomposing \mop{} flows and remapping work at runtime, rather than by building specialized task implementations. We study the effectiveness of this mechanism with two case studies: dynamic attention length and dynamic LoRA serving.



\boldpara{Dynamic Attention Length.}
We evaluate dynamic attention under four H100 workload regimes that vary both sequence length and batch structure, spanning short-context, long-context, and mixed-context cases (\autoref{fig:eval-end2end-uneven-context}). Mirage employs a fixed kernel template across these cases, and \sys{} outperforms it by a large margin in all regimes, with up to 6.18× lower latency. ThunderKittens adopts a host-side schedule and processor assignment from the input workload, but this adaptation is still built around a fixed specialized kernel. \sys{} consistently outperforms it by 15\% on average. The gap is especially pronounced in the short-context regime, where the overheads imposed by that fixed kernel structure are harder to amortize.

\boldpara{Dynamic LoRA serving.} 
LoRA~\cite{hu2022lora} specializes a shared base model with lightweight per-domain adapter weights. In serving, a single batch may mix requests that share the same adapter while others access different adapters. This creates highly variable effective matrix shapes, making fixed kernel schedules suboptimal.



We compare against sLoRA~\cite{sheng2024sloraservingthousandsconcurrent}, which uses stage-specific scheduling for the expand and shrink phases of LoRA execution. However, within each stage, the fixed kernel is largely agnostic to how requests are distributed across adapters. By contrast, \sys{} adapts the \mop{} flows of each adapter to the realized batch composition, improving SM occupancy and reducing makespan by up to 3.47x, as shown in \autoref{fig:eval-runtime-composability-throughput}.

\begin{figure}[t]
\centering
\includegraphics[width=\linewidth]{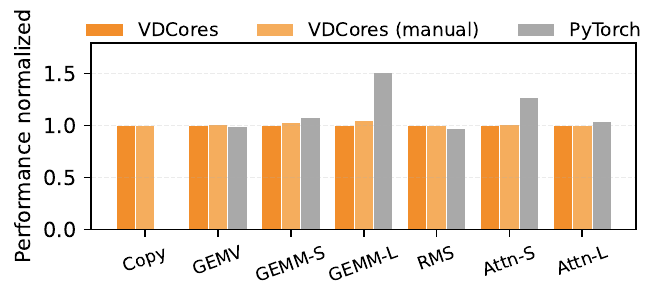}
\caption{Single-operator performance of \sys{} compared with \sys{} (manual), a manually tuned warp-specialization implementation, and with PyTorch.}
\label{fig:eval-runtime-single-kernel}
\end{figure}
\begin{figure}[!t]
\begin{minipage}[t]{0.48\columnwidth}
\vspace{0pt}
\centering
\includegraphics[width=\linewidth]{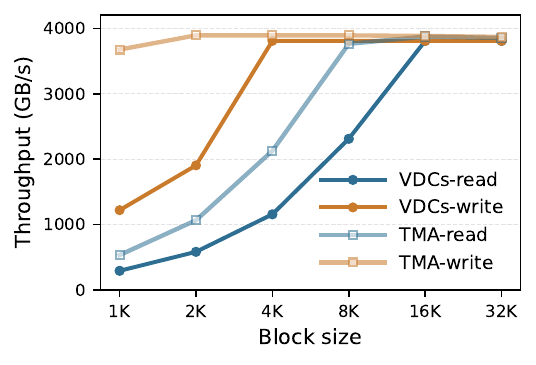}
\captionof{figure}{\sys{} (VDCs) memory operation efficiency at different block sizes.}
\label{fig:eval-runtime-overhead-breakdown}
\end{minipage}\hfill
\begin{minipage}[t]{0.47\columnwidth}
\vspace{0pt}
\centering
\scriptsize
\setlength{\tabcolsep}{2.6pt}
\renewcommand{\arraystretch}{1.03}
\begin{tabular}[t]{@{}p{0.68\linewidth}r@{}}
\toprule
Metric & Cycles \\
\midrule
Start up latency & 422 ns \\
\mop{} Initialization interval & 22ns \\
Ave memory load & 34ns \\
Ave memory st & 43ns \\
Ave memory ctrl & 22ns \\
\midrule
CUDA-core overhead & 3.1\% \\
\bottomrule
\end{tabular}
\captionof{figure}{Runtime latency metrics and aggregate runtime overhead. \footnotesize{Captured by NCU Compute Profiler.}}
\label{fig:eval-runtime-latency-metrics}
\end{minipage}
\end{figure}

\begin{figure}[!t]
\centering
    \centering
    \includegraphics[width=\linewidth]{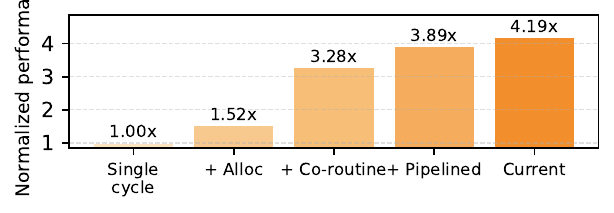}
    \captionof{figure}{Memory-core Optimization Breakdown.}
    \label{fig:eval-deepdive-tile-scheduler-memory-core}
\end{figure}
\subsection{Runtime Efficiency Study}
\label{sec:eval-ablation}

In this section, we focus on evaluating the design decisions in the virtual machine and executor by studying their performance implications.

\boldpara{Compute-bounded and memory-bounded kernels.}
We measure the runtime overhead of \sys{} on representative compute- and memory-intensive operators. We compare against VDE-manual, a hand-crafted warp-specialized implementation with similar pipelining but minimal overhead, and against PyTorch as a reference single-kernel baseline. As shown in \autoref{fig:eval-runtime-single-kernel}, \sys{} stays within 8\% of peak performance on average, while sustaining over 82\% of peak FLOPS and 93\% of peak memory bandwidth on H100.




\boldpara{Memory virtual core overhead.}
We sweep I/O block size and compare VDE’s effective bandwidth against the raw TMA issue rate to measure when virtual-core overhead is amortized. As \autoref{fig:eval-runtime-overhead-breakdown} shows, bandwidth approaches raw-instruction performance at 4KB for writes and 16KB for reads. Because real LLM operators typically move data at or above these granularities, the overhead of the virtual memory core is well amortized in practice. \autoref{fig:eval-runtime-latency-metrics} further breaks down the runtime cost over the lifetime of a virtual core and estimates that the aggregate overhead consumes only 3.1\% of total core time.






We ablate the key optimizations used to achieve this low overhead. We focus on instruction issue, because VMC throughput is largely determined by the control path that prepares and dispatches memory \mop{}s. As \autoref{fig:eval-deepdive-tile-scheduler-memory-core} shows, adding allocation support reduces latency from 366.9 to 242.2 cycles. Introducing coroutine execution further reduces latency to 112.0 cycles by improving overlap within the memory core, and pipelining lowers it again to 94.2 cycles. The final design reaches 87.6 cycles overall, yielding a 4.2× speedup over the naive implementation.




\section{Related Work}

\boldpara{Framework, Abstaction and Compiler for Asynchronous GPUs.}
CUTLASS, FlashInfer, and ThunderKittens provide optimized GPU libraries and templates that package expert implementation techniques such as tiling, pipelining, prefetching, and warp specialization \cite{cutlass, flashinfer, thunderkittens2024}. Flux and Mirage Persistent Kernel further optimize communication and persistent-kernel execution by structuring kernels into stages that overlap data movement and computation \cite{chang2024fluxfastsoftwarebasedcommunication, cheng2025mpk}. These frameworks make it easier to build high-performance GPU kernels, but they still express execution as specialized kernels with preplanned overlap and synchronization.

TVM, Relax, Fireiron, Triton, Cypress, and TileLang raise the abstraction level of GPU programming by generating optimized kernels from higher-level tensor, scheduling, or tile-based descriptions \cite{chen2018tvm, lai2025relax, hagedorn2020fireiron, tillet2019triton, cypress, tilelang}. TAWA \cite{chen2026tawa} further targets asynchronous GPU kernels by automatically applying warp specialization and pipeline scheduling. These systems reduce programmer effort and improve intra-kernel optimization, but their generated programs remain primarily kernel-centric: tiling, fusion, layout, and synchronization decisions are made before execution. VDCores differs by fundamental programming model shift: exposing memory and compute work as separate dependency-connected $\mu$ops, and results in assigning the coordination into a runtime layer.

\boldpara{Dataflow and Access-Execution Decoupled Architectures.}
Classic dataflow machines execute operations when their operands become ready, exposing parallelism through explicit dependencies rather than fixed instruction order \cite{dennis1975dataflow, arvind1986dataflow, arvind1990ttda, papadopoulos1990monsoon}. Decoupled Access/Execute architectures separate memory-access and compute streams to hide latency through producer--consumer queues \cite{smith1982dae, crago2011outrider, ham2015desc, chen2016datasupply, arnau2012mobilegpu, wang2017dac}. Many ML accelerators similarly use dataflow-style execution to make tensor movement, reuse, and pipeline structure explicit in hardware \cite{chen2014diannao, chen2014dadiannao, jouppi2017tpu, abts2020tsp}. These systems rely on specialized hardware, fixed dataflows, or general token-matching mechanisms. VDCores keeps the principles, but restricts them to GPU-friendly ML $\mu$ops and virtual flows and further use software cores to realize them on existing GPUs.

Compilers for dataflow, DAE, and spatial accelerators map programs into explicit graphs, streams, or hardware dataflows. They typically perform dependence analysis, tiling, placement, buffering, and scheduling to expose locality and parallelism to the target architecture. Ember applies compiler optimization for decoupled accelerator to irregular embedding operations in recommender models, improving performance and performance per watt \cite{siracusa2026ember}. VDCores converts the GPU to virutal decoupled hardware, further leveraging the existing compiling methods for global optimization.

\section{Discussion}
\label{sec:related}

\boldpara{Compiler support for \sys{}.}
We view \sys{} as a suitable abstraction layer for high-performance compilers.
At \mop{} layer,
Compilers could now generate efficient handler for single leveraging their static packing and local schedule optimization.
At system level, similar to dataflow compilers, compilers could be integrated into \mop{} generator and help with translating ML operators into \mop{} streams,
leveraging the \sys{} runtime harness of performance and correctness.
This split preserves compiler efficiency where it works best, while avoiding ad-hoc runtime tuning and leave them for \sys{}'s efficient runtime, also further support future learning-based or AI-assisted optimization.

\boldpara{Evolving Asynchronous GPU.}
New GPU generations continue to introduce new asynchronous resource domains beyond traditional CUDA-core execution. For example, recent architectures such as Blackwell expose new tensor-memory inside each SM, and modern GPUs increasingly provide distributed shared-memory capabilities across SM groups.

\sys{} can absorb these hardware changes without changing the decoupled model: we can either extend existing virtual-core implementations (e.g., VMC/VCC behaviors) or add new virtual-core types for the new resource domain. For instance, a additional specialized computation virtual cores can be introduced with keeping the same dependency-driven coordination.
On new virtual cores,
\sys{} runtime design could be reused, and it could integrate and overlap seamlessly into \sys{} system.




\boldpara{Supporting Heterogeneous Accelerators and Tiered Memory System.}
\sys{} shows good fit for emerging heterogeneous accelerators which explicitly decouple the hardware units for computation, memory movement and communication.
For example, in AWS Trainium \cite{bshara2024aws}, those jobs are handled by Neuron Core, DMA and CC-Core respectively.
\sys{} decoupled model
extends naturally beyond GPUs,
each resource domain is represented as a virtual core,
and cross-domain coordination is expressed through explicit dependency exchange rather
than implicit launch order.

When all backends implement the same decoupled interface and memory model,
\sys{} can also serve as a bridge abstraction across different heterogeneous hardware.
the compiler/runtime can compose end-to-end pipelines across dissimilar devices
without rewriting operator semantics for each target. Compared with ad-hoc
cross-device orchestration, this design preserves overlap opportunities while
maintaining a uniform integration boundary, similar to inside the single GPU device.
\section{Conclusion}
\label{sec:conclude}

We present \sys{}, a virtual decoupled-core programming framework and runtime for modern asynchronous GPUs. \sys{} decouples asynchronous resource units, enables auto-overlapping and dynamic-fusion optimizations for LLM serving, and enables composable optimization. We believe this design offers a practical path toward a unified, high-performance systems stack for next-generation LLM workloads.

\begin{small}
  \bibliographystyle{plain}
  \bibliography{all-defs,local,paper,mlsys}
\end{small}

\end{document}